\documentstyle[aps,preprint]{revtex}

\begin{document}
\title{Signatures of the excitonic memory effects in four-wave mixing processes in
cavity polaritons}
\author{Yu. P. Svirko and Makoto Kuwata-Gonokami\cite{boss}}
\address{Department of Applied Physics, Faculty of Engineering,\\
University of Tokyo and Cooperative Excitation Project of ERATO (JST),\\
7-3-1 Hongo, Bunkyo-ku, Tokyo 113-8656, Japan}
\date{\today}
\maketitle

\begin{abstract}
We report the signatures of the exciton correlation effects with finite
memory time in frequency domain degenerate four-wave mixing (DFWM) in
semiconductor microcavity. By utilizing the polarization selection rules, we
discriminate instantaneous, mean field interactions between excitons with
the same spins, long-living correlation due to the formation of biexciton
state by excitons with opposite spins, and short-memory correlation effects
in the continuum of unbound two-exciton states. The DFWM spectra give us the
relative contributions of these effects and the upper limit for the time of
the exciton-exciton correlation in the unbound two-exciton continuum. The
obtained results reveal the basis of the cavity polariton scattering model
for the DFWM processes in high-Q GaAs microcavity.
\end{abstract}


\narrowtext
Importance of the effects associated with the exciton-exciton interaction in
the nonlinear optical response at the fundamental band edge has been
revealed in a number of experiments in $\chi ^{(3)}$-regime. These effects,
which originate from the electromagnetic Coulomb interaction between
photo-generated carriers and Pauli exclusion principle, are referred to as
four particle correlation effects \cite{correlation0,correlation}. Extensive
effort has been devoted to investigate such correlations in semiconductors
by four-wave mixing spectroscopy, which gives us a unique opportunity to
evaluate limits of applicability of the fermionic mean field theory \cite
{fermion}. However, under actual experimental conditions, the nonlinear
optical measurements are affected by a number of intrinsic and extrinsic
effects, which are caused by non-electronic degrees of freedom. These
effects can not be discriminated in a nonlinear optical experiment in wide
spectral region, preventing the direct comparison between theory and
experiment.

However, the situation becomes more transparent when we restrict ourselves
to the spectral window close to the exciton resonance, where the effects of
four particle correlation can be classified in terms of spin dependent
interaction between excitons \cite{mukamel,sham,2}. In particular, a simple
model \cite{2}, referred to as the weakly interacting boson (WIB) model, in
which the interaction between $1s$ excitons is described by using two
parameters to account for the interaction between excitons with the same and
opposite spins, has allowed us to reproduce the overall features of the
polarization-sensitive DFWM spectra.

The remarkable efficiency of the WIB model in the describing the
polarization-sensitive DFWM measurements in time \cite{aoki} and frequency
domain \cite{2} has made it a promising and handy scheme to study the
correlation effects by methods of nonlinear spectroscopy. However, the
underlying physics and, especially, the role of second,- and higher-order in
exciton-exciton interactions effects \cite{inoue}, still remain unclear
because the model does not account for memory effects. These effects are
responsible for the temporal evolution of the $2\omega $-coherence within
the continuum of the unbound two-exciton states and the long-living
correlation due to the existence of biexciton, which has also been found to
be important in FWM at semiconductor band edge \cite{5,schaefer,9}. Note,
that the evolution equation for the excitonic polarization with account for
the biexciton state has been obtained in \cite{8} as a natural extension of
the WIB model. Similar semi-phenomenological treatment, which ignores the
memory effects due to the unbound two-exciton continuum, can be developed by
starting from the fermionic description of the excitonic nonlinearity \cite
{correlation,schaefer,kner}. The study of the effects of the bound and
unbound two-exciton states in the third-order optical response is especially
interesting in GaAs system, where - in strong contrast to the wide gap
semiconductors such as CuCl \cite{CuCl} or ZnSe \cite{6} - the biexciton
binding energy is the order of the exciton linewidth \cite{5}.

In order to elucidate the role of the memory effects at the fundamental band
edge, a special attention should be paid to the choice of the material for
the nonlinear-optical measurements. Specifically, the experiment should be
performed in a system where effects associated with the finite exciton
population and inhomogeneity are minimized. In this condition the effects of
the exciton population can be minimized by performing pump-probe experiment
at large detuning from the excitonic resonance. In this condition the
polarization-sensitive shift of the excitonic resonance, which is referred
to as the optical Stark effect, is the signature of the four-particle
correlation \cite{Stark}. At resonance condition, the exciton-cavity coupled
system is a well suited candidate to observe the signature of the
four-particle correlation. This is because strong coherent coupling between
exciton and photon in the high-Q microcavity, which leads to the formation
of cavity polaritons \cite{cavity} and suppresses incoherent effects in the
DFWM signal \cite{Shirane}. In this paper, we show that the study of
polarization-sensitive DFWM spectra in GaAs/AlGaAs QW embedded in the high-Q
microcavity allows us to discriminate the memory effects. By comparing the
results of the experiment and theory, we show that the correlation memory
time of excitons in GaAs is very short justifying the description of the
DFWM process, which is based on cavity polariton scattering. We also obtain
the relative contributions from the bound and unbound two-exciton states to
the nonlinear optical response.

The semiconductor microcavity investigated in this work has a single
12-nm-thick GaAs quantum well at the antinode of a $\lambda /2$-planar
microcavity, consisting of 22 and 14.5 pairs of distributed Bragg reflectors
for the bottom and topside respectively. All the measurements were performed
at 13 K. The linear reflection spectrum shows that the normal mode splitting
at zero detuning (4.3 meV) is larger than the linewidth of either exciton or
cavity mode (1.5 meV). The DFWM signal was measured in {\it (xxx)-}, {\it %
(xyy)-}, {\it (+++)}-, {\it (x++)-} and {\it (x+-)-}configurations, which
are abbreviated by the polarizations of the pump, test and signal beams,
respectively, in the self-pumped phase conjugation geometry \cite{2}. We
used the tunable picosecond pulses (pulse width of 1.9 psec and spectral
width of 0.7 meV) from a Kerr-lens mode-locked Ti:Sapphire laser with a 76
MHz repetition rate. In order to ensure that the measurements were performed
in the $\chi ^{(3)}$-regime, we examined excitation-power dependence of the
DFWM signal and found that it is proportional to $I_{pump}^{2}I_{test}$. The
detailed description of the experiment can be find in \cite{2}.

The DFWM spectra at zero exciton-cavity detuning are presented in Fig.\ref
{Fig1}a for different polarization configurations \cite{2}. In the {\it (x++)%
}- and {\it (+++)}-configurations, the spectra experiment show nearly the
same intensities for upper and lower polaritons. In the {\it (x+-)}-
configuration, where the DFWM signal is dominated by the cavity polariton
scattering is due to two-exciton states with zero angular momentum, the
signal at lower mode is found to be 2.5 times of that at upper mode. The
DFWM spectra in {\it (xxx)}- and {\it (xyy)}-configurations show a switching
between upper and lower modes.

In order to explain these results, we need to examine how the memory effects
in the exciton interaction manifest themselves in different polarization
configuration. Following \cite{9} we consider the resonant excitation only
and start from the equation of motion for the normalized complex amplitudes
of the right- and left-circular components of the excitonic polarization at
the frequency $\omega $, $p_{\pm }=<b_{\pm }>/(Vv_{e})^{1/2}$, where $b_{\pm
}$ is the exciton annihilation operator, $V$ and $v_{e}$ are the crystal and
exciton volume respectively, subscripts $"\pm "$ label right- and
left-circular components. The exciton volume is defined from the
conventional relationship between the exciton and interband dipole moments: $%
\mu _{ex}/\mu =(V/v_{e})^{1/2}$. The evolution equation for $p_{\pm }$ can
be presented in the following form: 
\begin{eqnarray}
&-i\frac{\partial p_{\pm }}{\partial t}+\Delta p_{\pm }=&(1-C|p_{\pm
}|^{2})\Omega _{\pm }+  \nonumber \\
&&-p_{\pm }^{\ast }\int_{0}^{\infty }F(\tau )p_{\pm }^{2}(t-\tau
)e^{-2i\Delta \tau }d\tau -p_{\mp }^{\ast }\int_{0}^{\infty }G(\tau
)p_{+}(t-\tau )p_{-}(t-\tau )e^{-2i\Delta \tau }d\tau  \label{p}
\end{eqnarray}
Here $\Omega _{\pm }=\mu E_{\pm }/\hbar $ are the Rabi frequencies, which
correspond to the right- and left-circular components of the electric field $%
E_{\pm }$ of the light wave at the QW, $\Delta =\omega _{e}-\omega -i\gamma $%
, $\omega _{e}$ and $\gamma $ are the exciton frequency and dephasing rate,
respectively; $C>0$ is the phase space filling{\bf \ (}PSF) constant \cite
{PSF}; $F(\tau )$ and $G(\tau )$ are memory functions, which account for
both instantaneous and retarded parts of the exciton-exciton interaction.

Since in our experiment, the pulse is longer than the exciton dephasing
time, we can consider the steady state approximation ignoring the time
dependence of slowly varying amplitudes $p_{\pm }$ and $\Omega _{\pm }$ in
Eq. (\ref{p}). By using Eq. (\ref{p}) and the evolution equation for the
electric field at the QW \cite{8,10}, the steady-state amplitudes of the
third-order polarization at the frequency $\omega $, which is responsible
for the DFWM signal in the phase conjugated geometry, can be presented in
the following form: 
\begin{equation}
p_{\pm }^{(3)}=A\{E_{\pm ,pump}^{2}[C\Delta +\int_{0}^{\infty }F(\tau
)e^{-2i\Delta \tau }d\tau ]+E_{\mp ,pump}^{2}\int_{0}^{\infty }G(\tau
)e^{-2i\Delta \tau }d\tau \}E_{\pm ,test}^{\ast }  \label{pp}
\end{equation}
where $A$ accounts for resonance enhancement of the electric field in the
microcavity, $E_{\pm ,pump}$ and $E_{\pm ,test}$ are amplitudes of the
electric field associated with the pump and test beams, respectively, at the
QW.

In order to show how the memory effects manifest themselves in the nonlinear
response, we separate the memory function $F(\tau )$ in terms of
instantaneous (mean field) part given by $\phi >0$, and retarded
(correlation) part \cite{9}, which is given by $\Phi (\tau )$: $F(\tau
)=\phi \delta (\tau )-\Phi (\tau )$. The mean field parameter $\phi $ is of
the first order in exciton-exciton interaction and describes the interaction
between excitons with same spins and zero center-of-mass momentum \cite
{3,hanamura}, while $\Phi (\tau )$ accounts for correlation effects of the
second- and higher-order of the interaction between two excitons. The memory
function $G(\tau )$ accounts for the effects arising from the interaction
between excitons with opposite spins. In this case, the first-order in
exciton-exciton interaction term vanishes and this memory function contains
correlation effects of the second- and higher order in interaction between
two excitons. In this configuration, there exists a bound state of two
excitons. Correspondingly, we separate $G(\tau )$ in terms of contribution
from the unbound two-exciton part given by $\Psi (\tau )$ and biexciton part
given by $\psi e^{i\omega _{B}\tau }$, where $\omega _{B}$ is the biexciton
binding energy: $G(\tau )=\Psi (\tau )-i\psi e^{i\omega _{B}\tau }$ \cite{9}.

In the high-Q microcavity, the strong coupling between excitons and photons
produces polariton modes, which dominate both linear and DFWM spectra. These
modes are referred to as lower and upper cavity polaritons. At zero
exciton-cavity detuning their frequencies are $\omega _{\alpha ,\beta
}=\omega _{e}\mp g$, respectively, where $g=(2\pi \omega \mu ^{2}/\hbar
nv_{e})^{1/2}$ is the energy of the dipole coupling between exciton and
photon and $n$ is the refractive index. The intensities of the
polarization-sensitive DFWM signal at frequencies of the lower and upper
polaritons for different polarization configurations can be obtained from (%
\ref{pp}) as follows: $I_{\alpha ,\beta }^{xxx}\propto |R+W\pm (Cg+\delta
R+\delta W)|^{2}$, $I_{\alpha ,\beta }^{xyy}\propto |R-W\pm (Cg+\delta
R-\delta W)|^{2}$, $I_{\alpha ,\beta }^{+++}\propto |R\pm (Cg+\delta R)|^{2}$
and $I_{\alpha ,\beta }^{x+-}\propto |W\pm \delta W|^{2}$, where 
\begin{eqnarray}
&&R=\phi -\int_{0}^{\infty }\Phi (\tau )e^{-2\gamma \tau }\cos 2g\tau d\tau 
\nonumber \\
&&\delta R=i\int_{0}^{\infty }\Phi (\tau )e^{-2\gamma \tau }\sin 2g\tau d\tau
\nonumber \\
&&W=\int_{0}^{\infty }\Psi (\tau )e^{-2\gamma \tau }\cos 2g\tau d\tau +\frac{%
(2ig+\omega _{B})\psi }{(2i\gamma +\omega _{B})^{2}-4g^{2}}  \nonumber \\
&&\delta W=i\int_{0}^{\infty }\Psi (\tau )e^{-2\gamma \tau }\sin 2g\tau
d\tau +\frac{2g\psi }{(2i\gamma +\omega _{B})^{2}-4g^{2}}  \label{w}
\end{eqnarray}

Normal mode splitting $2g$ is the major spectral characteristic of the
strongly coupled exciton-cavity system and, correspondingly, the role of the
memory effects in the excitonic nonlinear response is determined by its
ratio to the spectrum width of the memory functions. In order to clarify the
role of the memory effects in the DFWM spectra, we first examine the
long-time memory limit case, i.e. $2g>>\tau _{c}^{-1}$, where $\tau _{c}$ is
the correlation time of the memory functions $\Phi (\tau )$ and $\Psi (\tau
) $. By simplifying Eq. (\ref{w}) with account for $2g\tau _{c}>>1$ and $%
g>>\gamma ,\omega _{B}$ one can arrive at the following equations for the
polariton intensities: $I_{\alpha ,\beta }^{+++}\propto |\phi \pm Cg|^{2}$, $%
I_{\alpha ,\beta }^{xxx}\propto |\phi \pm (Cg-\psi /2g)|^{2}$ and $I_{\alpha
,\beta }^{xyy}\propto |\phi \pm (Cg+\psi /2g)|^{2}$. Since $\phi ,\psi >0$
and $\gamma ,\omega _{B}<2g$ one may see that $I_{\alpha }^{xyy}>I_{\alpha
}^{+++}$ (intensity of the lower polariton in {\it (xyy)}-configuration is
higher than that in {\it (}$+++${\it )}-configuration) and $I_{\beta
}^{xxx}>I_{\beta }^{+++}$ (intensity of the upper polariton in {\it (xxx)}%
-configuration is higher than that in {\it (}$+++${\it )}-configuration).
However, it can be clearly observed from the spectra 1n Fig. \ref{Fig1}a,
that such a conclusion contradicts to the experimental results making
assumption $2g\tau _{c}>>1$ is invalid.

Therefore, our experimental findings invoke the condition $2g<\tau _{c}^{-1}$%
. In such a case, we can substitute $e^{-2\gamma \tau }cos2g\tau \approx
1-2\gamma \tau -2g^{2}\tau ^{2}+...$ and $e^{-2\gamma \tau }sin2g\tau
\approx 2g\tau +...$, and neglect $\delta R$ in comparison with $Cg$.
Similarly, at $\omega _{B}<2g$ one can estimate $W\approx \int_{0}^{\infty
}\Psi (\tau )d\tau $ and $\delta W\approx -\psi /2g$. These gives $I_{\alpha
,\beta }^{xxx}\propto |R+W\pm (Cg+\delta W)|^{2}$, $I_{\alpha ,\beta
}^{xyy}\propto |R-W\pm (Cg-\delta W)|^{2}$, $I_{\alpha ,\beta }^{+++}\propto
|R\pm Cg|^{2}$ and $I_{\alpha ,\beta }^{x+-}\propto |W\pm \delta W|^{2}$
ensuring $I_{\alpha }^{+++}>I_{\alpha }^{xyy}>I_{\alpha }^{xxx}$ and $%
I_{\beta }^{+++}>I_{\beta }^{xxx}>I_{\beta }^{xyy}$. Note that since we
observe $I_{\alpha }^{x+-}>I_{\beta }^{x+-}$, the following relationship
holds: $W<\delta W<0$. The observed difference in the intensities of the
upper and lower polaritons in the {\it (x+-)}-configuration originates from
the bound two-exciton state. The ratio $I_{\alpha }^{x+-}/I_{\beta
}^{x+-}\approx 2.5$ obtained in the experiment is consistent with the
followed from the sum rule \cite{9} theoretical estimation $I_{\alpha
}^{x+-}/I_{\beta }^{x+-}\approx 1+2\omega _{B}/g$, for typical GaAs
biexciton binding energy \cite{5}. The calculated DFWM spectra with account
for the bound and unbound two-exciton states are presented in Fig. \ref{Fig1}%
b for $(-W):\delta W:R:Cg=0.7:0.23:0.12:1$. We would like to note here, that
our estimation $\tau _{c}<(2g)^{-1}$ is consistent with the results of the
calculation of the memory functions within the 1D-Hubbard model framework 
\cite{9}.

With account for $\delta R<<Cg$ and $|\delta W|<|W|$, the obtained result
returns the prediction of the polariton scattering model \cite{2}, which has
allowed us obtain Eq. (\ref{pp}) with the frequency independent $W$ and $R$.
In this model, these parameters account for the attraction and repulsion
between excitons with opposite and same spins, respectively, in the
phenomenological WIB Hamiltonian \cite{2,7}. The experimental spectra at
both zero (see Fig. \ref{Fig1}a) and arbitrary detuning for the above
mentioned polarization configurations have been explained by the following
relationship between the parameters of the polariton scattering model: $%
(-W):R:Cg=0.75:0.1:1$ \cite{2}. This has allowed us to conclude that the
attractive interaction between excitons and the PSF effect dominate in the
DFWM process in the high-Q microcavity. Apparently the polariton scattering
model failed to explain the difference in $I_{\alpha }^{x+-}$ and $I_{\beta
}^{x+-}$, which is due to the biexciton effect. Nevertheless, the
parameters, which has been obtained in \cite{2}, coincide with our present
estimations, because the biexcitonic effects do not affect significantly the
spectra in {\it (xxx)}- and {\it (xyy)}-configurations. The relatively small
value of the parameter $R$ obtained in the experiment is due to nearly
cancellation of the first- and higher-order in the exciton-exciton
interaction contributions \cite{inoue} to the resonant third-order
susceptibility. This cancellation has also been discussed in \cite{9} in
terms of constraints, which are imposed on the spectral density of the
memory functions by the sum rules.

In conclusion, we formulate the resonant DFWM results in terms of exciton
memory functions and show that the polarization-sensitive DFWM spectra give
us an important information on the memory effects in exciton-exciton
interaction. The intensity of the DFWM signal in the strongly coupled
exciton-cavity system is determined by both short-memory correlation in the
unbound two-exciton continuum and long-memory correlation associated with
the biexciton state. Both these effects give the second- and higher-order in
exciton-exciton interaction contributions to the resonance optical
nonlinearity, while the mean field, instantaneous contribution, which is of
the first-order in the exciton-exciton interaction, is nearly canceled. By
comparing the results of the experiment and theory in various polarization
configurations, we estimate the upper limit of the correlation time of the
memory functions, $\tau _{c}<<900$ fs, which describe the $2\omega $%
-coherence due to the continuum of the unbound two-exciton states. This also
allows us to show that the relative contribution to the resonant third-order
susceptibility from the biexciton in GaAs is about 30 percent. We show that
the short memory time of the exciton-exciton interaction permits to describe
the coherent optical response of the excitons in the high-Q semiconductor
microcavity in terms of the cavity polariton scattering model, which should
be extended to account for the biexciton state.

Discussions with Professor L. J. Sham, Professor A. Shimizu and Dr J. Inoue
are duly acknowledged. The microcavity has been provided by Professor H.
Sakaki and Dr T. Someya. We are also grateful to Dr C. Ramkumar, R. Shimano
and T. Aoki for helpful discussions. M.K.-G. is partially supported by a
grant-in-aid for COE Research from the Ministry of Education, Science,
Sports, and Culture of Japan.

\begin{figure}[tbp]
\caption{ Polarization-sensitive DFWM spectra. Left spectra: DFWM signal
intensity {\it (xyy)}-, {\it (xxx)} and {\it (+++)}-configurations shown by
solid, dashed and dotted lines, respectively. Right spectra: in {\it (x+-)}%
-and {\it (x++)}-configurations, shown by solid and dashed lines,
respectively, and scaled by factor 4. (a) experiment; (b) calculated with
account for the bound biexciton: $W:\protect\delta W:R:g\protect\nu
=.7:.23:.12:1$.}
\label{Fig1}
\end{figure}

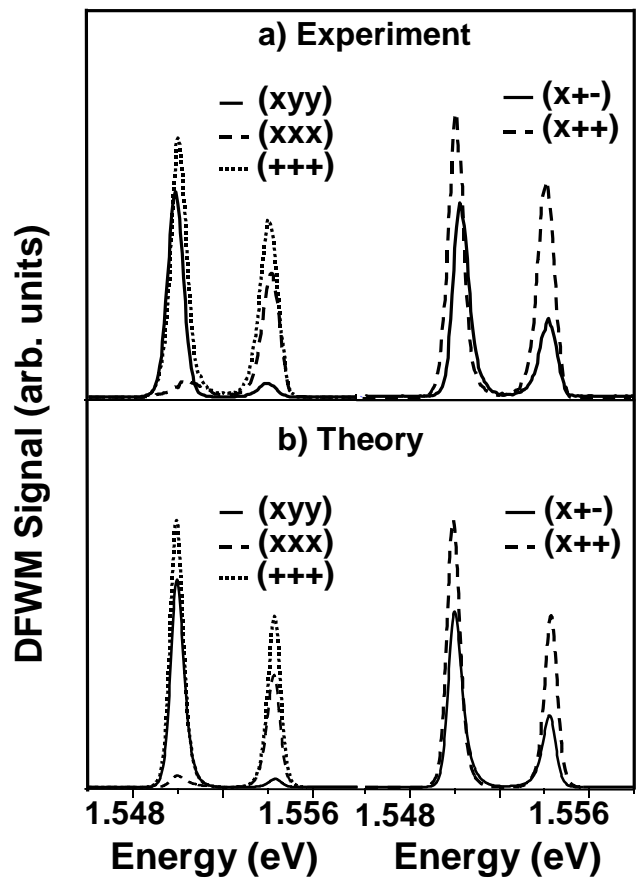
\end{document}